\documentclass[prd,aps,showpacs,twocolumn,twoside]{revtex4-1}
\usepackage{amsfonts,amsmath,amssymb,bm,graphicx}

\begin{document}

\title{Magnetic field instability in a neutron star driven by the electroweak electron-nucleon interaction versus the chiral magnetic effect}

\author{Maxim Dvornikov$^{a,b,c}$}
\email{maxim.dvornikov@usp.br}

\author{Victor B. Semikoz$^{b}$}
\email{semikoz@yandex.ru}

\affiliation{$^{a}$Institute of Physics, University of S\~{a}o Paulo, CP 66318, CEP 05314-970 S\~{a}o Paulo, SP, Brazil;
\\
$^{b}$Pushkov Institute of Terrestrial Magnetism, Ionosphere
and Radiowave Propagation (IZMIRAN), \\
142190 Troitsk, Moscow, Russia; \\
$^{c}$Physics Faculty, National Research Tomsk State University, \\
36 Lenin Ave., 634050 Tomsk, Russia}

\date{\today}

\begin{abstract}
We show that the Standard Model electroweak interaction of ultrarelativistic electrons with nucleons ($eN$ interaction) in a neutron star (NS) permeated by a seed large-scale helical magnetic field provides its growth up to 
$\gtrsim 10^{15}\thinspace\text{G}$ during a time comparable with the ages of young magnetars $\sim 10^4\thinspace\text{yr}$. The magnetic field instability  originates from the parity violation in the $eN$ interaction entering the generalized Dirac equation for right and left massless electrons in an external uniform magnetic field. We calculate the averaged electric current given by the solution of the modified Dirac equation containing an extra current for right and left electrons (positrons), which turns out to be directed along the magnetic field. Such current includes both a changing chiral imbalance of electrons and the $eN$ potential given by a constant neutron density in NS. Then we derive the system of the kinetic equations for the chiral imbalance and the magnetic helicity which accounts for the $eN$ interaction. By solving this system, we show that a sizable chiral imbalance arising in a neutron protostar due to the Urca-process $e^-_\mathrm{L} + p\to N + \nu_\mathrm{eL}$  diminishes very rapidly because of a huge chirality flip rate. Thus the $eN$ term prevails the chiral effect providing a huge growth of the magnetic helicity and the helical  magnetic field.
\end{abstract}

\pacs{97.60.Jd, 11.15.Yc, 95.30.Qd}



\maketitle


Some neutron stars, called magnetars, having magnetic fields $B\sim 10^{15}-10^{16}\thinspace\text{G}$, can be considered as strongest magnets in our universe~\cite{Mer08}. Despite the existence of various models for the generation of such strong fields, based, e.g., on the turbulent dynamo~\cite{Duncan}, the origin of magnetic fields in magnetars is still an open problem. Recently, in Ref.~\cite{Yamamoto2} the authors tried to apply the chiral magnetic effect~\cite{Vilenkin,Nielsen}, adapted successfully for the QCD plasma~\cite{Kharzeev}, to tackle the problem of magnetic fields in magnetars. The approach of Ref.~\cite{Yamamoto2} implies the chiral kinetic theory, where Vlasov equation is modified when adding the Berry curvature term to the Lorentz force~\cite{Yamamoto1}.

The fate of such a chiral plasma instability is based on the Adler anomaly in QED with the nonconservation of the pseudovector current for massless fermions $\bar{\psi}\gamma_{\mu}\gamma_5\psi$ in external electromagnetic fields. Since this current is the difference of right $j_{\mu}^\mathrm{R}$ and left $j_{\mu}^\mathrm{L}$ currents, the {\it assumption of a seed imbalance} between their densities given by the difference of chemical potentials, $(n_\mathrm{R} -  n_\mathrm{L})\sim \mu_5=(\mu_\mathrm{R} - \mu_\mathrm{L})/2\neq 0$, where $n_\mathrm{R,L}$ are the densities of right and left fermions (electrons) and $\mu_\mathrm{R,L}$ are their chemical potentials, could lead to the magnetic field instability we study here adding electroweak interactions in the Standard Model (SM).

The same effect (while without weak interactions) was used in Ref.~\cite{Boyarsky:2011uy} to study the self-consistent evolution of the magnetic helicity in the hot plasma of the early Universe driven by the change of the lepton asymmetry $\sim \mu_5$. In Ref.~\cite{Boyarsky:2011uy} it was showed that such an asymmetry diminishes, $\mu_5\to 0$, due to the growth of the chirality flip rate in the cooling universe through electron-electron ($ee$) collisions, $\Gamma_f \sim \alpha_\mathrm{em}^2 \left( m_e/3T \right)^2$, where $\alpha_\mathrm{em} = e^2/4\pi \approx 1/137$ is the fine structure constant, $m_e$ is the electron mass, and $T$ is the plasma temperature.

This negative result encouraged the appearance of Ref.~\cite{Boyarsky:2012ex}, where another mechanism for the generation of magnetic fields was proposed. It is based on the parity violation in electroweak plasma resulting in the nonzero Chern-Simons (CS) term $\Pi_2$ that enters the antisymmetric part of the  photon polarization operator in plasma of massless particles. Here we adopt the notation for the CS term from Ref.~\cite{Boyarsky:2012ex}. In Ref.~\cite{Dvornikov:2013bca}, a similar CS term $\Pi_2^{(\nu l)}$, based on the neutrino interactions with charged leptons, was calculated. Basing on this calculation, the magnetic field instability driven by neutrino asymmetries was revealed. This instability is implemented in different media such as the hot plasma of the early universe and a supernova (SN) with a seed magnetic field.

The amplification of a seed magnetic field during the SN burst driven by a non-zero electron neutrino asymmetry $\Delta n_{\nu_e}\neq 0$ which enters the CS term $\Pi_2^{(\nu e)}$ was suggested in Ref.~\cite{Dvornikov:2013bca} to explain the generation of strongest magnetic fields in magnetars. Note that after the SN burst a cooling neutron star (NS) as the corresponding SN remnant emits equally neutrinos and antineutrinos. Thus, the neutrino asymmetry vanishes. The inclusion of the electroweak $ee$-interaction with a stable fraction of degenerate electrons $n_e \approx \text{const}$ instead of the $\nu e$ interaction with vanishing neutrino asymmetry $\Delta n_{\nu_e}\to 0$ has no sense since the corresponding parity violating CS term $\Pi_2^{(ee)}$ tends to zero in the static limit $\omega\to 0$ for an electron gas, $\Pi_2^{(ee)}\to 0$, as found in Ref.~\cite{Dvornikov2}.

In the present work we suggest to take into account the electroweak electron-nucleon ($eN$) interaction providing a long time acting source of the magnetic field instability
that plays a role of a CS term in the pseudovector electron current ${\bf J}_5 = \Pi_2^{(eN)}{\bf B}$. Instead of the Matsubara technique used in Refs.~\cite{Dvornikov:2013bca,Dvornikov2}, here we calculate the total electric current in SM (additive to the standard ohmic current) solving the Dirac equation for the massless right and left electrons (positrons) in a magnetic field.
%

We start the derivation of the aforementioned CS term with solving the Dirac equation for a massless electron in the magnetic field $\mathbf{B}= (0,0,B)$ accounting for the electroweak $eN$ interaction in NS. This equation reads as
\begin{equation}\label{Dirac}
  \left[
    \gamma^{\mu}
    \left(
      \mathrm{i}\partial_{\mu}+eA_{\mu}
    \right) -
    \gamma^{0}
    \left(
      V_{\mathrm{L}} P_{\mathrm{L}} + V_{\mathrm{R}} P_{\mathrm{R}}
    \right)
  \right]\psi_e=0,
\end{equation}
%
%
where $\gamma^{\mu}=\left(\gamma^{0},\bm{\gamma}\right)$ are the
Dirac matrices, $A^{\mu}=\left(0,0,Bx,0\right)$ is the vector potential, $P_{\mathrm{L,R}}=(1\mp\gamma^{5})/2$
are the chiral projection operators, $\gamma^{5}=\mathrm{i}\gamma^{0}\gamma^{1}\gamma^{2}\gamma^{3}$, and $e>0$ is the absolute value of the electron charge. 

In Eq.~\eqref{Dirac} we assume that there are no macroscopic fluid (nucleon) currents in NS. The effective potentials $V_{\mathrm{L,R}}$ in Eq.~\eqref{Dirac} are given by the SM Lagrangian of the $eN$ interaction via neutral currents in the Fermi approximation (see, e.g., Ref.~\cite{Oku82}),
\begin{align}\label{lagrangian}
  \mathcal{L} = & \sqrt{2} G_\mathrm{F}
  \bar{\psi}_e\gamma_{\mu}
  \left(
    g_\mathrm{L}^{(e)} P_\mathrm{L} + g_\mathrm{R}^{(e)}P_\mathrm{R}
  \right)
  \psi_e
  \notag
  \\
  & \times
  \left[
    \bar{\psi}_n\gamma^{\mu} \psi_n -
    (1 - 4\xi) \bar{\psi}_p\gamma^{\mu} \psi_p
  \right],
\end{align}
where $G_\mathrm{F} \approx 1.17 \times 10^{-5}\thinspace\text{GeV}^{-2}$ is the Fermi constant
$g_\mathrm{L}^{(e)} = - 1/2 + \xi$ and $g_\mathrm{R}^{(e)}=\xi$  are the standard coupling constants in SM with the Weinberg parameter $\xi = \sin^2\theta_W \approx 0.23$, and $\psi_{n,p}$ are the neutron and proton wave functions. We reduced  the total $eN$ Lagrangian in Ref.~\cite{Oku82} to Eq.~\eqref{lagrangian} omitting the axial nucleon currents $\sim \bar{\psi}_{n,p} \gamma^\mu \gamma^5 \psi_{n,p}$ irrelevant to our problem.

Taking the statistical averaging $\langle \dots \rangle$ in Eq.~\eqref{lagrangian} over the equilibrium (Fermi) distributions of nucleons in a neutron star and recalling that macroscopic nucleon currents are absent, i.e. $\langle \bar{\psi}_{n,p} \bm{\gamma} \psi_{n,p} \rangle = 0$, we get the following definition of $V_\mathrm{R,L}$ to be used in Eq.~(\ref{Dirac}):
\begin{align}\label{VLR}
  V_{\mathrm{L}} = &
  - \frac{G_\mathrm{F}}{\sqrt{2}}
  \left[
    n_n - n_p(1 - 4\xi)
  \right](2\xi - 1),
  \notag
  \\
  V_{\mathrm{R}} = &
  - \frac{G_\mathrm{F}}{\sqrt{2}}
  \left[
    n_n - n_p(1 - 4\xi)
  \right]
  2\xi,
\end{align}
where $n_{n,p} = \langle \psi_{n,p}^\dag \psi_{n,p} \rangle$ are the number densities of neutrons and protons.


Let us decompose $\psi_e$ in the chiral projections as $\psi_e=\psi_{\mathrm{L}}+\psi_{\mathrm{R}}$,
where $\psi_{\mathrm{L,R}}=P_{\mathrm{L,R}}\psi_e$. Then, using Eq.~(\ref{Dirac})
we get that $\psi_{\mathrm{L,R}} = e^{-\mathrm{i}E_{\mathrm{L,R}}t+\mathrm{i}p_{y}y+\mathrm{i}p_{z}z} \psi_{\mathrm{L,R}}(x)$, where

\begin{align}\label{eq:psiL}
\psi_{\mathrm{L,R}}^{(\mathrm{n})}(x)= & \frac{1}{4\pi\sqrt{E_{\mathrm{L,R}}-V_{\mathrm{L,R}}}}
\notag
\\
& \times
\left(\begin{array}{c}
\sqrt{E_{\mathrm{L,R}}-V_{\mathrm{L,R}}\mp p_{z}}u_{\mathrm{n}-1}\\
\mp\mathrm{i}\sqrt{E_{\mathrm{L,R}}-V_{\mathrm{L,R}}\pm p_{z}}u_{\mathrm{n}}\\
\mp\sqrt{E_{\mathrm{L,R}}-V_{\mathrm{L,R}}\mp p_{z}}u_{\mathrm{n}-1}\\
\mathrm{i}\sqrt{E_{\mathrm{L,R}}-V_{\mathrm{L,R}}\pm p_{z}}u_{\mathrm{n}}
\end{array}\right),
\notag\\
\psi_{\mathrm{L,R}}^{(0)}(x)= & \frac{1}{2\pi\sqrt{2}}\left(\begin{array}{c}
0\\
u_{0}\\
0\\
\mp u_{0}
\end{array}\right).
\end{align}
Here $\psi^{(\mathrm{n})}_{\mathrm{L,R}}$ corresponds to $\mathrm{n}=1,2,\dotsc$, $\psi^{(0)}_{\mathrm{L,R}}$ to $\mathrm{n}=0$, $\eta=\sqrt{eB}x+p_{y}/\sqrt{eB}$, $u_{\mathrm{n}}(\eta) = \left( eB / \pi \right)^{1/4} \exp(-\eta^{2}/2) H_{\mathrm{n}}(\eta)/\sqrt{2^{\mathrm{n}}\mathrm{n}!}$, and
$H_{\mathrm{n}}(\eta)$ is the Hermite polynomial. The upper signs in Eq.~\eqref{eq:psiL} stay for $\psi_{\mathrm{L}}$ and the lower ones for $\psi_{\mathrm{R}}$. To derive Eq.~\eqref{eq:psiL} we use the $\gamma$ matrices in the Dirac representation as in Ref.~\cite{ItzZub80}. The energy levels $E_{\mathrm{L,R}}$ in Eq.~(\ref{eq:psiL})
can be obtained from the following expression:
\begin{equation}\label{energyLR}
  \left(E_{\mathrm{L,R}}-V_{\mathrm{L,R}}\right)^{2} =
  p_{z}^{2}+2eB\mathrm{n}.
\end{equation}
The normalization factors in Eq.~\eqref{eq:psiL} correspond to
\begin{multline}
  \int
  \left(
    \psi_\mathrm{L,R}
  \right)_{\mathrm{n}p_{y}p_{z}}^{\dagger}
  \left(
    \psi_\mathrm{L,R}
  \right)_{\mathrm{n}'p'_{y}p'_{z}}\mathrm{d}^{3}x
  \\
  =
  \delta_{\mathrm{n}\mathrm{n}'}
  \delta
  \left(
    p_{y}-p'_{y}
  \right)
  \delta
  \left(
    p_{z}-p'_{z}
  \right),
\end{multline}
since the chiral projections $\psi_\mathrm{L,R}$ are independent. It is worth mentioning that a more general solution of Eq.~\eqref{Dirac}, which accounts for the nonzero electron mass, was found in Ref.~\cite{BalPopStu11}.


The spinors in Eq.~\eqref{eq:psiL} are then used to calculate the induced electric current which has a nonzero projection on the $z$ axis $\sim \bar{\psi}_e\gamma^3\psi_e$. Analogously to Ref.~\cite{Vilenkin} one shows that the averaged current gets the contribution from the main Landau level $\mathrm{n}=0$ only.  It should be noted that massless particles have a strong correlation between their momentum and helicity. Thus, at $\mathrm{n}=0$, left electrons have $p_z > 0$, whereas right ones have $p_z < 0$.

Making the statistical averaging with the Fermi-Dirac distribution  of left and right electrons (positrons) $f_{e,\bar{e}}(E)=\left[\exp(\beta ( E \mp \mu_\mathrm{L,R})+1\right]^{-1}$, where $\beta=1/T$ is the reciprocal temperature, $\mu_\mathrm{L,R}$ are their chemical potentials, and the lower sign stays for positrons, then using Eq.~\eqref{energyLR}, one obtains the component of the current $J_z$ along the magnetic field in the form,
\begin{align}\label{eq:j5gen}
  J_z = &
  \frac{e^{2} B}{4\pi^{2}}
  \bigg\{
    \int_{-\infty}^{0}\mathrm{d}p_z
    \left[
      f_e
      \left(
        -p_z+V_{\mathrm{R}}
      \right) -
      f_{\bar{e}}
      \left(
        -p_z-V_{\mathrm{R}}
      \right)
    \right]
      \notag
      \\
      & -
    \int_{0}^{+\infty}\mathrm{d}p_z
    \left[
      f_e
      \left(
        p_z+V_{\mathrm{L}}
      \right) -
      f_{\bar{e}}
      \left(
        p_z-V_{\mathrm{L}}
      \right)
    \right]
  \bigg\}.
\end{align}
Basing on Eq.~\eqref{eq:j5gen} and introducing vector notations, we derive the averaged induced current in the final form as
\begin{equation}\label{current}
  {\bf J}= \frac{2\alpha_\mathrm{em}}{\pi}(\mu_5 + V_5){\bf B},
\end{equation}
which is additive to the ohmic current ${\bf J}_\text{Ohm}$ in a standard QED plasma. It should be noted that Eq.~\eqref{current} is valid for any electron temperature.

The current in Eq.~\eqref{current} is proportional to $\alpha_\mathrm{em}$ 
and consists of the two parts:
the vector term  given in QED by the {\it pseudoscalar} coefficient $\mu_5=(\mu_\mathrm{R} - \mu_\mathrm{L})/2$ ($\mu_5\to - \mu_5$ under spatial inversion) and the pseudovector current ${\bf J}_5=(2\alpha_\mathrm{em}/\pi)V_5{\bf B}=\Pi_2^{(eN)}{\bf B}$ given in SM  by the {\it scalar} factor $V_5=(V_{\mathrm{L}} - V_{\mathrm{R}})/2$.
Indeed, after the statistical avaraging the interaction Lagrangian in Eq.~\eqref{lagrangian} becomes,
\begin{equation}\label{Ltransf}
  \mathcal{L} =
  \frac{1}{2} (V_\mathrm{L} + V_\mathrm{R})
  \bar{\psi}_e\gamma_0\psi_e +
  \frac{1}{2} (V_\mathrm{R} - V_\mathrm{L})
  \bar{\psi}_e\gamma_0\gamma_5\psi_e.
\end{equation}
The factor $\bar{\psi}_e\gamma_0\gamma_5\psi_e$, in the parity violation term of Eq.~\eqref{Ltransf}, is the pseudoscalar with respect to the spatial inversion $P=P^{-1}=\gamma_0$, since $P\gamma_0\gamma_5P^{-1}= - \gamma_0\gamma_5$. Thus $V_5$ should be {\it scalar}; cf. Ref.~\cite{DS1}. The true pseudoscalar both for P-inversion and Lorentz transformation should be $\bar{\psi}\gamma_5\psi$. It should be noted that one looses the Lorentz invariance in a medium with the selected reference frame like NS at rest.

The weak interaction coefficient in Eq.~\eqref{current}
\begin{equation}\label{f5}
  V_5 =
  \frac{G_\mathrm{F}}{2\sqrt{2}}
  [n_n - (1 - 4\xi)n_p],
\end{equation}
is of the order $V_5 \approx G_\mathrm{F} n_n/2\sqrt{2}  = 6\thinspace\text{eV}$ in NS with $n_n = 1.8 \times 10^{38} \thinspace \text{cm}^{-3}$, which corresponds to $\rho_n=3\times 10^{14} \thinspace \text{g} \cdot \text{cm}^{-3}$, since $n_p \ll n_n$. On the first sight, the electromagnetic QED term in the current in Eq.~(\ref{current}), $\sim \mu_5$, seems to be much bigger than the weak one in Eq.~(\ref{f5})~\cite{fnYambigmu5}. However, we show below that the latter term remains as a stable source of the magnetic field instability in NS while the former one vanishes, $\mu_5\to 0$, e.g., for helical magnetic fields with the maximum helicity contrary to the statement in Ref.~\cite{Yamamoto1} that an imbalance $\mu_5\neq 0$ could lead to the generation of strong magnetic fields in magnetars.


The evolution of the magnetic field in the presence of the induced current in Eq.~(\ref{current}), proportional to the magnetic field, obeys the modified Faraday equation; cf. Ref.~\cite{Dvornikov:2013bca}. However, it is more convenient to study the evolution of the magnetic helicity density
\begin{equation}\label{heldef}
  h(t) = \frac{1}{V} \int \mathrm{d}^3 x (\mathbf{A} \cdot \mathbf{B}),
\end{equation}
where $\mathbf{A}$ is the 3D vector potential and $V$ is the volume of space. Defining the helicity density spectrum $h(k,t)$ as $h(t)=\smallint \mathrm{d}k h(k,t)$ and accounting for the induced current in Eq.~(\ref{current}), which includes both the chiral imbalance contribution $\sim \mu_5$ and the electroweak term $\sim V_5$, we get the kinetic equation for $h(k,t)$
which is the generalization of Eq.~(6) in Ref.~\cite{Boyarsky:2011uy},
\begin{align}\label{helicityevolution}
  \frac{\partial h(k,t)}{\partial t} = &
  - \frac{2k^2h(k,t)}{\sigma_\mathrm{cond}}
  \notag
  \\
  & +
  \frac{\alpha_\mathrm{em}}{\pi}
  \left[
    \frac{k(\Delta \mu + 2V_5)}{\sigma_\mathrm{cond}}
  \right]h(k,t).
\end{align}
Here $\Delta \mu=\mu_\mathrm{R}- \mu_\mathrm{L}=2\mu_5$ and we have just assumed as in Ref.~\cite{Boyarsky:2011uy} the maximal helicity field configuration, i.e. the magnetic energy density reads $\rho_\mathrm{B}(t)=\smallint \mathrm{d} k \rho_\mathrm{B}(k,t)= (1/2) \smallint\mathrm{d}k k h(k,t)$. It is worth to be mentioned that the sign of the $\Delta \mu$ term in Eq.~\eqref{helicityevolution} is opposite to that in Ref.~\cite{Boyarsky:2011uy} since we use the different definition of $\gamma^5$.

Then we should derive the kinetic equation which governs the chiral imbalance evolution, which is complementary to Eq.~\eqref{helicityevolution}. Using Eq.~\eqref{heldef} and the Maxwell equations, we get the helicity density change in the standard form,
\begin{equation}\label{helevol}
  \frac{\mathrm{d}h(t)}{\mathrm{d}t}=
  -\frac{2}{V}\int \mathrm{d}^3x ({\bf E}\cdot{\bf B}).
\end{equation}
where ${\bf E}$ is the electric field. Then, accounting for the Adler anomaly for the pseudovector current in electromagnetic fields, $\partial_{\mu}(j^{\mu}_\mathrm{R} - j_\mathrm{L}^{\mu}) = \partial_{\mu}(\bar{\psi}\gamma^{\mu}\gamma^5\psi) = (2\alpha_\mathrm{em}/\pi) ({\bf E}\cdot{\bf B})$,
we derive the conservation law involving $h(t)$ and $n_\mathrm{R,L}$,
\begin{equation}\label{law}
  \frac{{\rm d}}{{\rm d}t}
  \left[
    n_\mathrm{R} - n_\mathrm{L} + \frac{\alpha_\mathrm{em}}{\pi}h(t)
  \right]=0,
\end{equation}
which is valid in a QED plasma.
%

Taking into account that $n_\mathrm{L,R}=\mu_\mathrm{L,R}^3/3\pi^2$ and assuming that $\mu_\mathrm{L}\sim \mu_\mathrm{R}\sim \mu$, where $\mu$ is the chemical potential of the degenerate electron gas in NS, that is true at least at the beginning of the imbalance in NS, we get that
$
n_\mathrm{R} - n_\mathrm{L}
\approx 2\mu_5\mu^2/\pi^2.
$
Eventually we obtain from Eq.~(\ref{law}), using the expression for $\partial h(k,t)/\partial t$ in Eq.~\eqref{helicityevolution}, the evolution equation for $\mu_5$,
\begin{align}\label{kinetics2}
  \frac{\mathrm{d}\mu_5}{\mathrm{d}t}= &
  \frac{\pi\alpha_\mathrm{em}}{\mu^2 \sigma_\mathrm{cond}}
  \int \mathrm{d} k \thinspace k^2h(k,t)
  \notag
  \\
  & -
  \left[
    \frac{2\alpha_\mathrm{em}^2\rho_\mathrm{B}(t)}{\mu^2\sigma_\mathrm{cond}}
  \right]
  (\mu_5 + V_5) - \Gamma_f\mu_5.
\end{align}
In Eq.~\eqref{kinetics2} we added the rate of chirality-flip processes, $\Gamma_f\simeq (m_e/\mu)^2\nu_\mathrm{coll}$, given by the Rutherford electron-proton ($ep$) collision frequency $\nu_\mathrm{coll}=\omega_p^2/\sigma_\mathrm{cond}$ without flip. Here $\omega_p = \mu \sqrt{4\alpha_\mathrm{em}/3\pi}$  is the plasma frequency in a degenerate ultrarelativistic electron gas and $\sigma_\mathrm{cond}$ is the electric conductivity in a degenerate electron-proton plasma consisting of ultrarelativistic degenerate electrons and non-relativistic degenerate protons~\cite{fneecoll}. Note that in a degenerate electron gas $\nu_\mathrm{coll}$ depends on the temperature $T$; cf. Ref.~\cite{Kelly}. This is due to the Pauli principle when all electron states with the momenta $0\leq p\leq \mu$ are busy, i.e. $ep$ scattering is impossible at $T=0$.



One can see that Eq.~(\ref{kinetics2}) is different from the simplified kinetic approach $\mathrm{d}\mu_5/\mathrm{d}t = \Gamma_\mathrm{inst}\mu_5 - \Gamma_f\mu_5$, where $\Gamma_\mathrm{inst} = \alpha_\mathrm{em}^2 \mu_5$ is a factor providing the magnetic field growth, used in Refs.~\cite{Yamamoto1,Yamamoto2}. The first term in the rhs of
Eq.~(\ref{kinetics2}) can be really estimated as $\sim \alpha_\mathrm{em}^2\mu_5^2$ for all ``equal'' parameters $\mu\sim \mu_5\sim \sigma_\mathrm{cond}$ that is not the case we rely on. The more important difference is the appearance of the second term $\sim \rho_\mathrm{B}$ that is the back reaction from the magnetic field that diminishes an imbalance $\mu_5$.

Let us choose the simplest case of the monochromatic helicity density spectrum $h(k,t)=h(t)\delta (k - k_0)$ where  we can vary
the wave number $k_0$ and the magnetic field scale $\Lambda_\mathrm{B} = k_0^{-1}$ to find later some critical regimes for the imbalance evolution $\mu_5(t)$ through Eq. (\ref{kinetics2}). Using the dimensionless functions $\mathcal{M}(\tau)= (\alpha_\mathrm{em}/\pi k_0) \mu_5(t)$ and $\mathcal{H}(\tau)= (\alpha_\mathrm{em}^2/2k_0\mu^2) h(t)$ which depend on the dimensionless diffusion time $\tau = (2k_0^2 / \sigma_\mathrm{cond}) t$ we can recast the self-consistent system of Eqs.~(\ref{helicityevolution}) and~(\ref{kinetics2}) as
\begin{align}\label{system}
  \frac{\mathrm{d}\mathcal{M}}{\mathrm{d}\tau} = &
  (1 - \mathcal{M} - \mathcal{V})\mathcal{H} - \mathcal{G}\mathcal{M},
  \nonumber
  \\
  \frac{\mathrm{d}\mathcal{H}}{\mathrm{d}\tau} = &
  - (1 - \mathcal{M} - \mathcal{V})\mathcal{H}.
\end{align}
Here for fixed $V_5=6\thinspace\text{eV}$ the dimensionless parameters $\mathcal{V} = (\alpha_\mathrm{em} / \pi k_0) V_5$ and
$\mathcal{G} = (\sigma_\mathrm{cond} / 2k_0^2) \Gamma_f= (2\alpha_\mathrm{em} / 3\pi) \left( m_e / k_0 \right)^2$ are the function of the parameter $k_0$ only. Note that $\mathcal{G}$ does not depend on the conductivity $\sigma_\mathrm{cond}$ since the rate of the chirality flip can be estimated as
$\Gamma_f\simeq \left( m_e/\mu \right)^2 \nu^{(\text{no flip})}_\mathrm{coll}$ where in the magnetohydrodynamic plasma $\nu^{(\text{no flip})}_\mathrm{coll} = \omega_p^2 / \sigma_\mathrm{cond}$ is the $ep$ collision frequency without flip. The dimensionless diffusion time $\tau$ depends on the conductivity found in Ref.~\cite{Kelly}
\begin{equation}\label{conductivity}
  \sigma_\mathrm{cond} =
  \frac{1.6\times 10^{28}}{(T/10^8\thinspace\text{K})^2}
  \left(
    \frac{n_e}{10^{36}\thinspace\text{cm}^{-3}}
  \right)^{3/2}\thinspace\text{s}^{-1},
\end{equation}
that is valid for cooling NS matter consisting of degenerate non-relativistic nucleons and ultrarelativistic degenerate electrons.
%

For the magnetic field scale $\Lambda_\mathrm{B}$ comparable with the NS radius $R_\mathrm{NS}=10\thinspace\text{km}$, or for the small wave number $k_0=1/R_\mathrm{NS}=2\times 10^{-11}\thinspace\text{eV}$, one gets the electroweak interaction contribution in Eq.~(\ref{system}) as
$\mathcal{V}
=7\times 10^8$
coming from the current in Eq.~\eqref{current},
where we substitute the small $V_5=6\thinspace\text{eV}$.
We choose the initial chiral imbalance as $\mu_5(0)=1\thinspace\text{MeV}\ll \mu$, where for $n_e=\mu^3/3\pi^2=10^{36}\thinspace\text{cm}^{-3}$ in Eq. (\ref{conductivity}) the electron chemical potential equals to $\mu=60\thinspace\text{MeV}$. Hence at the beginning the dimensionless chiral imbalance $\mathcal{M}(0)= (\alpha_\mathrm{em} / \pi k_0) \mu_5(0)\simeq 10^{14}$ is much bigger than the electroweak term $\mathcal{V}$. On the first glance, such inequality could be expected comparing electromagnetic and weak interaction effects, $\mathcal{M}(0) \gg \mathcal{V}=\text{const}$. We assume also the constant temperature in a cooling NS $T=10^8\thinspace\mathrm{K}$~\cite{fnTconst}. Therefore the electric conductivity  in Eq.~(\ref{conductivity}) is also constant, $\sigma_\mathrm{cond}=10^7\thinspace\text{MeV}$.

The dimensionless chirality flip rate
\begin{equation}\label{flip}
  \mathcal{G}=\frac{2\alpha_\mathrm{em}}{3\pi}\left(\frac{m_e}{k_0}\right)^2=10^{30},
\end{equation}
is huge for the given small $k_0=2\times 10^{-14}\thinspace\text{keV}$. If we change $m_e=511\thinspace\text{keV} \to m_\text{eff}=\mu \sqrt{\alpha_\mathrm{em}/2\pi}$~\cite{Braaten:1993jw}, the rate in Eq.~(\ref{flip}) would be even bigger diminishing $\mu_5$ faster in the first line in Eq.~(\ref{system}). Finally, for the acceptable initial magnetic field $B_0=10^{12}\thinspace\text{G}$, the initial helicity density
$h(0)= B_0^2/k_0=
2\times 10^{13}\thinspace\text{MeV}^3$ gives $\mathcal{H}(0)=
(\alpha_\mathrm{em}^2/2k_0\mu^2) h(0) = 6\times 10^{21}$.

\begin{figure}
  \includegraphics[scale=.4]{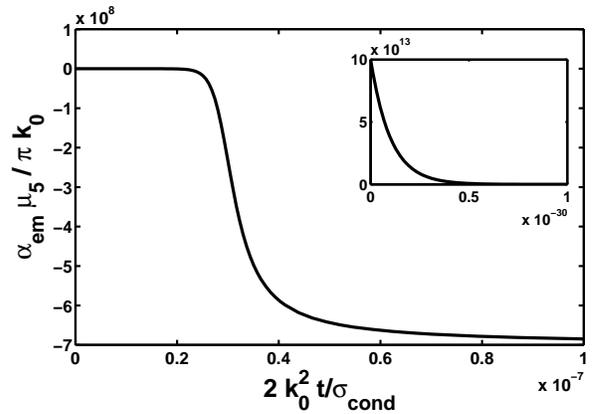}
  \caption{The dimensionless chiral imbalance $\mathcal{M}$ versus $\tau$. The horizontal axis of the
  main plot starts at $\tau \gtrsim 10^{-30}$. The inset shows the evolution of $\mathcal{M}$ in the initial time interval corresponding to $\tau < 10^{-30}$.
  \label{fig:mu5}}
\end{figure}

\begin{figure}
  \includegraphics[scale=.4]{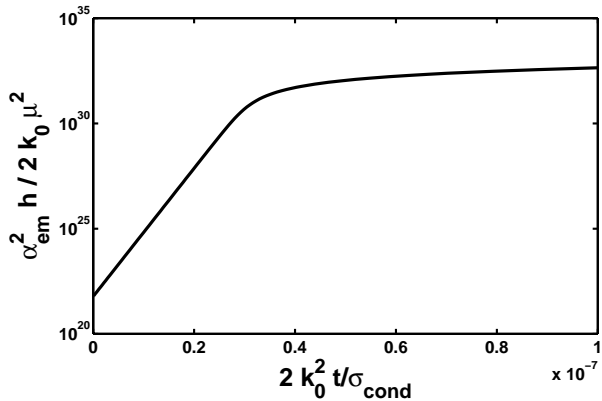}
  \caption{The dimensionless helicity $\mathcal{H}$ versus $\tau$.
  \label{fig:helicity}}
\end{figure}

We solved the system of the self-consistent kinetic equations in Eq.~(\ref{system}) numerically for the adopted $\mathcal{V}$ and $\mathcal{G}$ as well as the initial conditions $\mathcal{M}(0)=10^{14}$ and $\mathcal{H}(0)=6\times 10^{21}$ chosen above. In Fig.~\ref{fig:mu5} we plot the evolution of the chiral imbalance $\mathcal{M}(\tau)$.
In the inset, one can see how a large chirality imbalance $\mu_5 \sim \mathcal{O}(\text{MeV})$ vanishes owing to the huge chirality flip rate in Eq.~(\ref{flip}), $\mu_5 \to 0$, during a very short time $\tau \sim 10^{-30}$ corresponding to $t\sim 10^{-12}\thinspace\text{s}$. In the main plot one finds a sharp slope for $\mathcal{M}$ somewhere at $\tau\approx 3\times 10^{-8}$ that corresponds to the time $t\sim 8000\thinspace\text{yr}$. The obtained critical time is of the order of young magnetar ages~\cite{Mer08}.
In Fig.~\ref{fig:helicity} we see that, at the same time $\tau\approx 3\times 10^{-8}$, the magnetic helicity density $\mathcal{H}$ grows on about ten orders of magnitude,
that corresponds to the growth of $B=\sqrt{k_0 h}$ on the five orders of magnitude, just getting $B\simeq 10^{17}\thinspace\text{G}$ if we started from the seed field $B_0=10^{12}\thinspace\text{G}$.

It is interesting to mention that, in Fig.~\ref{fig:mu5}, a {\it positive} primeval chirality imbalance, $\mu_5=(\mu_\mathrm{R}-\mu_\mathrm{L})/2 >0$, which appears, e.g., due to the direct Urca process, $e_\mathrm{L}^- + p\to n + \nu_\mathrm{eL}$, becomes {\it negative}, $\mu_\mathrm{R}-\mu_\mathrm{L} <0$. This happens due the simultaneous growth of the helicity density $h$ (see in Fig.~\ref{fig:helicity}) that amplifies the negative derivative $\mathrm{d}\mathcal{M}/\mathrm{d}\tau <0$ much more intensively than the chirality flip $\sim \mathcal{G}$. Vice versa, the attenuation of $\mathcal{M}$ owing to the chirality flip is more important at the first stage illustrated in the inset of Fig.~\ref{fig:mu5}. Since $\mathcal{M}\to - \mathcal{V}= - 7\times 10^8$ ($\mu_5\to - 6\thinspace\text{eV}$, see in Fig.~\ref{fig:mu5}), while the decreasing sum $\mathcal{V} + \mathcal{M}$ remains positive, the value of the positive derivative $\mathrm{d}\mathcal{H}/\mathrm{d}\tau > 0$ diminishes, or the helicity evolution simulates a saturation, see in Eq.~(\ref{system}) and in Fig.~\ref{fig:helicity}.

Finally we notice that rather helical magnetic fields determine the evolution of the  chiral imbalance $\mu_5(t)$ than a non-zero seed $\mu_5\neq 0$ leads to the growth of the magnetic helicity density $h=B^2/k_0$ or the magnetic field itself. This imbalance starting from a sizable value $\mu_5\sim \mathcal{O}(\text{MeV})$ decreases down to the $eN$ interaction term $|\mu_5| \sim V_5\sim 6\thinspace\text{eV}$. We stress that namely the electroweak interaction term $V_5 > |\mu_5|$ drives the amplification of the seed magnetic field in NS, see in the second line in Eq.~(\ref{system}). If one takes into account the cooling of a neutron star, $\mathrm{d}T/\mathrm{d}t <0$, a more realistic model to generate strong magnetic fields in magnetars can be developed. We plan to do that in our future work.

Of course, we considered here only the largest scale $k_0^{-1}=R_\mathrm{NS}=10\thinspace\text{km}$ as the most interesting case for magnetic fields in NS.
Our model is simplified both due to the choice of the maximum helicity density $kh(k,t)=2\rho_\mathrm{B}(k,t)$ instead of the more general inequality $kh(k,t)\leq 2\rho_\mathrm{B}(k,t)$ \cite{Biskamp}, and owing to the choice of the monochromatic helicity density spectrum $h(k,t)=h(t)\delta (k-k_0)$. The generalization of our model, e.g.,  accounting for an initially non-helical magnetic field, the continuous magnetic energy spectrum, complicates the problem. This requires to solve the system
of kinetic equations for the magnetic helicity density and magnetic energy density instead of
the single  Eq. (\ref{helicityevolution}) here. We skip also the stage of a supernova collapse with non-equilibrium
processes on that time, considering in our model mostly long time intervals $\sim (10^3 - 10^4)\thinspace\text{yr}$
for a thermally relaxed NS core.

We would like to mention that recently, in Ref.~\cite{grabow}, the application of the chiral plasma instability in SN was also criticized because the chirality flip was underestimated in Ref.~\cite{Yamamoto2} in the approximation $\left( m_e/\mu \right)^2\ll 1$. Instead of a tedious calculation made in Ref.~\cite{grabow}, we can reproduce in a simpler way the flip rate $\Gamma_f$ obtained by the authors in Ref.~\cite{Yamamoto2} and demonstrate why their derivation is invalid. Indeed, in Ref.~\cite{Yamamoto2} the authors incorrectly relied on the flip rate $\Gamma_f\sim \alpha_\mathrm{em}^2 \left( m_e/\mu \right)^2\mu_5$ meaning rather that the collision frequency without flip, entering the flip rate as $\Gamma_f=(m_e/\mu)^2\nu_\mathrm{coll}^{(\text{no flip})}$,  is given by the common formula $\nu_\mathrm{coll}^{(\text{no flip})}=\sigma n_e\simeq \left( \alpha_\mathrm{em}^2/\mu^2 \right) \mu^3=\alpha_{em}^2\mu_5$. Here it was assumed that $\mu\sim \mu_5$, using the electron density $n_e\sim \mu^3$ and the Rutherford cross-section for $ep$ collisions $\sigma\sim \alpha_\mathrm{em}^2/\langle E\rangle^2$, where $\langle E\rangle \sim \mu$ is the mean electron energy. Such estimate of the flip rate $\Gamma_f$ is incorrect for a degenerate electron gas because the Pauli principle was not taken into account.

To resume we have suggested here a novel mechanism for the magnetic field amplification in NS based on the $eN$ electroweak interaction. For this purpose, in Eq.~\eqref{current}, we have generalized the CS term, derived in Ref.~\cite{Vilenkin}, to include the electroweak interaction of right and left ultrarelativistic degenerate electrons with nucleons. Then, in Eqs.~\eqref{helicityevolution} and~\eqref{kinetics2}, we have obtained the new system of kinetic equations for the evolution of the chiral imbalance and the magnetic helicity. This system accounts for the $eN$ interaction and the back reaction. Finally we have applied our results to predict the magnetic field growth in NS up to values observed in magnetars.

It should be noted that our model is absolutely different
from the well-known approach put forward in Ref.~\cite{Duncan} based on a strong turbulent convection in the core of SN and the
fast dynamo operating only for a few seconds, being driven by the high neutrino luminosity $L_{\nu} > 10^{52}\thinspace\text{erg}\cdot\text{s}^{-1}$ at that time. It should be noted that, in Ref.~\cite{VinKui06}, it was found that protostars, which were progenitors to some magnetars, did not seem to reveal a fast rotation as required in Ref.~\cite{Duncan}. We also refute
the arguments in Ref.~\cite{Yamamoto2} suggested to explain the generation of strong magnetic fields in magnetars based on the chiral magnetic instability.

We acknowledge L.~B.~Leinson and D.~E.~Kharzeev for fruitful discussions as well as S.~B.~Popov for communications. M.D. is thankful to FAPESP (Brazil) for the grant No. 2011/50309-2, to the Competitiveness Improvement Program and the Academic
D.~I.~Mendeleev Fund Program at the Tomsk State University for a partial support, as well as
to RFBR (research project No. 15-02-00293) for a partial support.


\begin{thebibliography}{100}

\bibitem{Mer08}
  S.~Mereghetti,
  The strongest cosmic magnets:
  Soft Gamma-ray Repeaters and Anomalous X-ray Pulsars,
  Astron. Astrophys. Rev. \textbf{15}, 225 (2008).

\bibitem{Duncan}
  R.~C.~Duncan and C. Thompson,
  Formation of very strongly magnetized neutron stars - Implications for gamma-ray bursts,
  Asprophys. J. \textbf{392}, L9 (1992).

\bibitem{Yamamoto2}
  A.~Ohnishi and N.~Yamamoto,
  Magnetars and the chiral plasma instabilities,
  arXiv:1402.4760.

\bibitem{Vilenkin}
  A. Vilenkin,
  Equilibrium parity-violating current in a magnetic field,
  Phys. Rev. D \textbf{22}, 3080 (1980).

\bibitem{Nielsen}
  H.~B.~Nielsen and M.~Ninomiya,
  Adler-Bell-Jackiw anomaly and Weyl fermions in crystal,
  Phys. Lett. B \textbf{130}, 389 (1983).

\bibitem{Kharzeev}
  K.~Fukushima, D.~E.~Kharzeev, and H.~J.~Warringa,
  The chiral magnetic effect,
  Phys. Rev. D \textbf{78}, 074033 (2008).

\bibitem{Yamamoto1}
  Y.~Akamatsu and N.~Yamamoto,
  Chiral plasma instabilities,
  Phys. Rev. Lett. \textbf{111}, 052002 (2013).

\bibitem{Boyarsky:2011uy}
  A.~Boyarsky, J.~Fr\"{o}hlich, and O.~Ruchayskiy,
  Self-consistent evolution of magnetic fields and chiral asymmetry in the early Universe,
  Phys. Rev. Lett. \textbf{108}, 031301 (2012).

\bibitem{Boyarsky:2012ex}
  A.~Boyarsky, O.~Ruchayskiy, and M.~Shaposhnikov,
  Long-range magnetic fields in the ground state of the Standard Model plasma,
  Phys. Rev. Lett. \textbf{109}, 111602 (2012).

\bibitem{Dvornikov:2013bca}
  M.~Dvornikov and V.~B.~Semikoz,
  Instability of magnetic fields in electroweak plasma driven by neutrino asymmetries,
  J. Cosmol. Astropart. Phys. 05 (2014) 002.

\bibitem{Dvornikov2}
  M. Dvornikov,
  Impossibility of the strong magnetic fields generation in an electron-positron plasma,
  Phys. Rev. D \textbf{90}, 041702 (2014).


\bibitem{Oku82}	
  L.~B.~Okun,
  \textit{Leptons and Quarks}
  (North-Holland Publishing Company, Amsterdam, 1982),
  pp.~209--211.

\bibitem{ItzZub80}
  C.~Itzykson and J.-B.~Zuber,
  \textit{Quantum Field Theory}
  (McGraw-Hill, New York, 1980), pp.~691--696.

\bibitem{BalPopStu11}
   I.~A.~Balantsev, Yu.~V.~Popov, and A.~I.~Studenikin,
   On the problem of relativistic particles motion in a strong magnetic field and dense matter,
   J. Phys. A \textbf{44}, 255301 (2011).

\bibitem{DS1}
  M.~Dvornikov and V.~B.~Semikoz,
  Lepton asymmetry growth in the symmetric phase of an electroweak
  plasma with hypermagnetic fields versus its washing out by sphalerons,
  Phys. Rev. D \textbf{87}, 025023 (2013).


\bibitem{fnYambigmu5}
  For instance, in Ref.~\cite{Yamamoto1} it was assumed that $\mu_5 \sim 200\thinspace\text{MeV}$, as expected for the chiral asymmetry close to the Fermi surface, $\mu_5\sim \mu$.

\bibitem{Braaten:1993jw}
  E.~Braaten and D.~Segel,
  Neutrino energy loss from the plasma process at all temperatures and densities,
  Phys. Rev. D \textbf{48}, 1478 (1993).
		
\bibitem{fneecoll}
  The effects of both $ee$ collisions and the scattering of electrons by the neutron magnetic moment are minor. See the comments on this issue in Ref.~\cite{Kelly}.

\bibitem{Kelly}
  D.~C.~Kelly,
  Electrical and thermal conductivities of a relativistic degenerate plasma,
  Astrophys. J. \textbf{179}, 599 (1973).

\bibitem{fnTconst}
  Of course, the temperature diminishes during cooling of a neutron star, primarily, due to the neutrino emission. We know how to manage that in our scenario varying the conductivity in Eq.~(\ref{conductivity}). Nevertheless, here to match competing mechanisms, the chiral magnetic instability and the novel one caused by the electroweak $eN$ interaction, we assume for simplicity that $\sigma_\mathrm{cond}=\text{const}$.

	
\bibitem{Biskamp}	
  D. Biskamp,
  \textit{Magnetohydrodynamic Turbulence}
  (Cambridge University Press, Cambridge, 2003).

\bibitem{grabow}
  D.~Grabowska, D.~Kaplan, and S.~Reddy,
  The role of the electron mass in damping chiral magnetic instability in
  supernova and neutron stars,
  arXiv:1409.3602.

\bibitem{VinKui06}
  J.~Vink and L.~Kuiper,
  Supernova remnant energetics and magnetars:
  no evidence in favour of millisecond proto-neutron stars,
  Mon. Not. R. Astron. Soc. \textbf{370}, L14 (2006).

\end{thebibliography}
\end{document}